\documentclass[lettersize,journal]{IEEEtran}

\usepackage{xcolor}
\usepackage{cite}
\usepackage{amsmath,amssymb,amsfonts,amsbsy}
\usepackage{algorithmic}
\usepackage{graphicx}
\usepackage{textcomp}
\DeclareMathOperator*{\argmin}{arg\, min\, }
\def\BibTeX{{\rm B\kern-.05em{\sc i\kern-.025em b}\kern-.08em
    T\kern-.1667em\lower.7ex\hbox{E}\kern-.125emX}}
\def\BibTeX{{\rm B\kern-.05em{\sc i\kern-.025em b}\kern-.08em
    T\kern-.1667em\lower.7ex\hbox{E}\kern-.125emX}}
\usepackage{lineno}
\usepackage{subcaption}
\usepackage[normalem]{ulem}

\newcommand{\red}[1]{\textcolor{black}{#1}}

\begin{document}

\def\ManuscriptType{ARXIV}
\def\IEEEpaper{IEEE}
\def\ARXIVpaper{ARXIV}

\title{Extended Depth-of-Field Lensless Imaging\\using an Optimized Radial Mask}

\author{Jos\'{e} Reinaldo Cunha Santos A. V. Silva Neto, Tomoya Nakamura, Yasushi Makihara,\\and Yasushi Yagi,~\IEEEmembership{Senior Member, IEEE},
\thanks{This work was supported by JST FOREST under Grant JPMJFR206K. (Corresponding author: Tomoya Nakamura.)

José Reinaldo Cunha Santos A. V. Silva Neto, 
Tomoya Nakamura, 
Yasushi Makihara, 
and Yasushi Yagi 
are with the SANKEN, Osaka University, Osaka, 567-0047, Japan (e-mail: vieira@am.sanken.osaka-u.ac.jp; nakamura@am.sanken.osaka-u.ac.jp; makihara@am.sanken.osaka-u.ac.jp; 
yagi@am.sanken.osaka-u.ac.jp).

This paper has supplementary downloadable material available at http://ieeexplore.ieee.org., provided by the author. The material includes additional experimental results from both simulations and using our developed prototype camera.
}}

\ifx\ManuscriptType\IEEEpaper
\markboth{IEEE transactions on computational imaging, vol. xx, YEAR XXXX}{C. S. A. V. Silva Neto \MakeLowercase{\textit{et al.}}: Extended Depth-of-Field Lensless Imaging using an Optimized Radial Mask}
\fi

\maketitle

\begin{abstract}
The freedom of design of coded masks used by mask-based lensless cameras is an advantage these systems have when compared to lens-based ones. We leverage this freedom of design to propose a shape-preserving optimization scheme for a radial-type amplitude coded mask. \red{Due to the depth-independency of the radial mask's point spread function, they can be used for extending the effective depth of field (DOF) of a lensless imaging system. In this paper we optimized a coded mask for improved frequency response, while retaining its radial characteristics and therefore extended-DOF capabilities.} We show that our optimized radial mask achieved better overall frequency response when compared to naive implementations of a radial mask. We also quantitatively and qualitatively demonstrated the extended DOF imaging achieved by our optimized radial mask in simulations by comparing it to different non-radial coded masks. Finally, we built a prototype camera to validate the extended DOF capabilities of our coded mask in real scenarios.
\end{abstract}


\section{Introduction}


\begin{figure*}[!t]
    \centering
    \includegraphics[width=\linewidth]{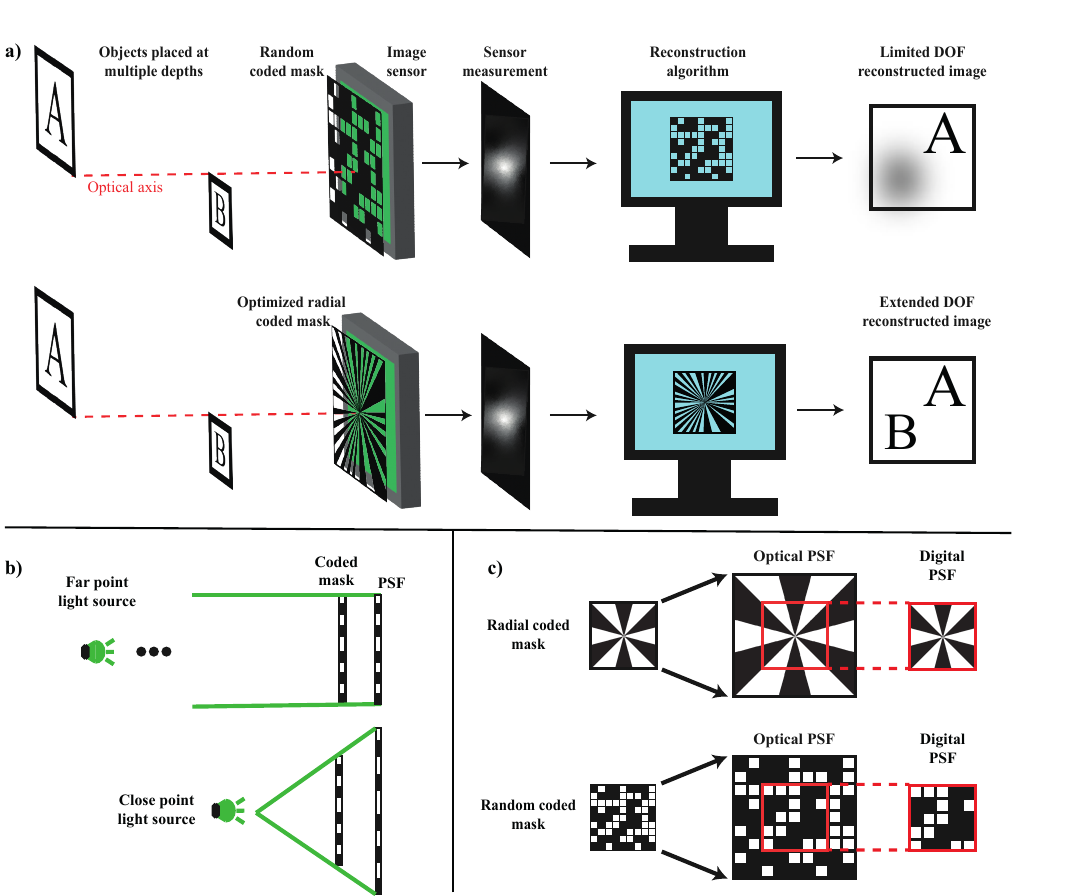}
    \caption{(a)~Complete pipeline for a mask-based lensless imaging system, from the acquisition of the image sensor measurements to the reconstruction through a computational algorithm. The pipeline on top exemplifies the limitation of a non-radial coded mask at reconstructing objects at very different distances from the camera. On the bottom pipeline, we illustrate the extended DOF achieved by a radial type coded mask. (b)~Schematic diagram of the depth dependency of a PSF, as it radially scales up when a light point source becomes closer to a coded mask. 
    (c)~Concept on the scaling-invariant property of a radial mask, as its digital PSF observed by an image sensor (red square) remains unchanged even when the optical PSF becomes large. This does not hold true for a non-radial mask such as a random mask, as its observed digital PSF (red square) changes.}
    \label{fig:intro_complete}
\end{figure*}

\ifx\ManuscriptType\IEEEpaper
\IEEEPARstart{M}{ask-based} lensless cameras are formed by replacing the lenses from a traditional camera with a coded mask~\cite{Boominathan2022}. This replacement makes it possible to produce smaller, lighter, and cheaper cameras when compared to traditional lens-based ones. The most notable difference between these two types of cameras is that the coded mask of a lensless camera multiplexes the incoming light, mapping one point in the ambient scene to many pixels in the image sensor, thus encoding the ambient scene information in visually uninformative sensor measurements.
\else 
Mask-based lensless cameras are formed by replacing the lenses from a traditional camera with a coded mask~\cite{Boominathan2022}. This replacement makes it possible to produce smaller, lighter, and cheaper cameras when compared to traditional lens-based ones. The most notable difference between these two types of cameras is that the coded mask of a lensless camera multiplexes the incoming light, mapping one point in the ambient scene to many pixels in the image sensor, thus encoding the ambient scene information in visually uninformative sensor measurements.
\fi

The sensor measurement formation process is often modeled as a two-dimensional~(2D) convolution between the point spread function~(PSF), which is an intensity impulse response in the spatial domain, of an optical system involving a coded mask and the incoming light from the scene~\cite{Goodman1996, Antipa2018,Asif2017}, as
\begin{equation}
    \textbf{b} = crop[\textbf{h}*\textbf{v}] = \textbf{A}\textbf{v},
    \label{eq:conv}
\end{equation}
where 
$\textbf{b} \in \mathbb{R}^{N_\mathrm{b} \times 1}$ is a vector representing the captured sensor measurements,
$crop[\cdot]$ is the cropping function of an image defined by the effective area of the pixel of an image sensor,
$\textbf{h} \in \mathbb{R}^{N_\mathrm{v} \times 1}$ is a vector representing a PSF, 
$*$ is the 2D convolution operator, and 
$\textbf{v} \in \mathbb{R}^{N_\mathrm{v} \times 1}$ is a vector representing the spatial light intensity of a scene.
$N_\mathrm{b} \in \mathbb{N}$ and $N_\mathrm{v}\in \mathbb{N}$ are the pixel count of captured and scene images, respectively.
We also define a matrix~$\textbf{A} \in \mathbb{R}^{N_\mathrm{b} \times N_\mathrm{v}}$ to simply denote the operations of 2D convolution followed by cropping. The model of Eq.~\eqref{eq:conv} assumes a shift-invariant system as an approximation~\cite{Martz1962}.

Ordinarily, the sensor measurements do not contain apparent visual information, and in order to reconstruct the scene information $\textbf{v}$, a reconstruction algorithm is employed.
In general, the regularized error-minimization method with iterative algorithms, which is stable against noise, is often employed as the image reconstruction method.
The minimization problem is defined as:
\begin{equation}
    \tilde{\textbf{v}} = 
    \operatorname*{argmin}_{\textbf{v}\geq 0}
    \| \textbf{b} - \textbf{A}\textbf{v} \|^2_2 
    + \tau \Psi(\textbf{v}),
\end{equation}
where $\Tilde{\textbf{v}} \in \mathbb{R}^{N_\mathrm{v} \times 1}$ is a vector representing a reconstructed image,
$\| \cdot \|_2$ is $\ell_2$-norm of a vector, 
$\tau$ is a constant value for controlling the effectiveness of the regularization,
$\Psi(\cdot)$ is a regularizer, which is often a combination of linear transformation and $\ell_1$-norm calculation.
Figure~\ref{fig:intro_complete}(a) illustrates the complete pipeline of a mask-based lensless camera. 

\subsection{Depth of Field in Lensless Imaging}
Similarly to traditional lens-based cameras, mask-based lensless imagers have a limited depth of field~(DOF), especially when shooting scenes that include a large distance range~\cite{Tan2017}. 
For a lensless system, this limitation stems from the fact that the PSF as a shadow of a coded mask is depth dependent. As illustrated in Fig.~\ref{fig:intro_complete}(b), object planes at different distances produce different PSFs. 
Considering this, the generalized forward model for imaging 3D scene is modeled as follows:
\begin{equation}
    \textbf{b} = crop\left[\sum_{z} \textbf{h}_z*\textbf{v}_z\right] = \sum_z\textbf{A}_z\textbf{v}_z,
    \label{eq:3Dforward}
\end{equation}
where $z > 0$ is the distance of the object plane measured from the mask plane, and $\sum_z$ is the incoherent summation of the depth-dependent information.
$\textbf{h}_z$ and $\textbf{v}_z$ are $z$-dependent vectors of $\textbf{h}$ and $\textbf{v}$, respectively, and $\textbf{A}_z$ is a $z$-dependent matrix of $\textbf{A}$.
Note that the effect of occlusion by opaque objects is ignored for simplicity.
The reconstruction process using a single PSF is as follows:
\begin{equation}
    \tilde{\textbf{v}}_{z{^*}} = 
    \operatorname*{argmin}_{\textbf{v}\geq 0}
    \| \textbf{b} - \textbf{A}_{z{^*}}\textbf{v} \|^2_2 
    + \tau \Psi(\textbf{v}),
    \label{eq:3Dinv}
\end{equation}
where $z^*$ is the distance value to be used for reconstruction, $\textbf{A}_{z{^*}} \in \mathbb{R}^{N_\mathrm{b} \times N_\mathrm{v}}$ is a matrix representing a forward operator corresponding to $z^*$, and $\tilde{\textbf{v}}_{z{^*}} \in \mathbb{R}^{N_\mathrm{v} \times 1}$ is a vector representing the reconstructed image with using $\textbf{A}_{z{^*}}$.
As indicated in Eq.~\eqref{eq:3Dinv}, only the depth range that matches an assumed depth in the reconstruction process $
z{^*}$ can be reconstructed correctly.
In other words, the gap between an object's real depth and the depth used for calibration of the PSF incurs a worse reconstruction.
This limitation on the correctly-reconstructable distance range with a single reconstruction filter can be defined as a DOF of the lensless imaging system.
This property is sometimes utilized in digital refocusing~\cite{Shimano2018}, depth-map acquisition~\cite{Zheng2020}, 3D imaging~\cite{Antipa2018,Adams2017, Tian2022}; however, it is undesirable in applications such as 2D image analysis because it impairs the spatial characteristics of the image.

To address this issue, the simplest way is to set the distance between the sensor and the mask closer. This method can extend the DOF in the same way as existing lens cameras; however, the angular resolution of the imaging system is sacrificed because the angle of view is also extended at the same time. 
Another simple method is to measure the PSFs at multiple distances in advance, reconstruct multiple images using all of the PSFs, and merge only the areas in focus from the resulting images into a single image~\cite{Hua2023}. This method is often used for lens-based imaging as called focal-stack photography~\cite{Kutulakos2009,Huang2022}. However, the challenges of lensless imaging are that it requires numerous reconstruction processes and contrast calculation processes, which are computationally expensive.
Furthermore, its synthesis accuracy depends on the complexity of the texture and structure of a scene.
A more advanced method is to use compressive holography~\cite{Brady2009} to reconstruct a 3D tomographic image~\cite{Antipa2018} and then generate an all-in-focus 2D image just by axial integration. Compressive holography reconstructs a sparse 3D tomographic image from a 2D image using a compressed sensing framework~\cite{Candes2008}.
This method is very interesting as a new method for tomography, but its computational cost is high because it needs to solve an ill-posed problem.
In addition, its accuracy severely depends on the sparsity of the scene.

To extend the DOF in lensless imaging while avoiding the above problems, several methods have been proposed that take advantage of the design freedom of a mask in lensless imaging.
For example, Hua {\it et al.} proposed a method to realize an approximately depth-invariant PSF by sweeping the mask design during a single exposure, which, combined with post-imaging deconvolution, extends the DOF~\cite{Hua2020}.
This method called {\it SweepCam} can be said to be an application of the focal-sweep method~\cite{Kuthirummal2011} to lensless cameras.
This is a reasonable method to obtain an extended DOF image only with fast and stable computation, but it requires the use of a refreshable-type mask such as a liquid-crystal spatial light modulator~(SLM) and sacrifices temporal resolution.
As a method that can be realized with a fixed mask, Gill proposed the use of an odd-symmetry phase grating~\cite{Gill2013OL}. A lensless camera constructed using this mask has an extended DOF when compared to one using a normal mask~\cite{Gill2013}; However, one limitation is that the depth-invariant region of the PSF is limited to near the center of the mask, where incoming light of point-symmetric locations of the ambient scene are enforced to have their phases destructively interfere with each other at the image plane. As a result, the degree of depth-invariance is limited in principle.

On the other hand, in a past study, we proposed the radial coded mask whose PSF is the depth invariant in the whole area at the image plane~\cite{nakamura_etal_radial_IAOC_2020}. 
A radial mask is an amplitude-transmission mask that has structure only in the rotational direction and no structure in the radial direction.
Such a mask is usually used to measure optical transfer functions~\cite{Goodman1996} or to generate a non-diffracting beam~\cite{Rasouli2018}.
Figure~\ref{fig:intro_complete}(c) illustrates the concept of the depth-invariant PSF by a radial mask.
From a geometry-based point of view, the change of a PSF due to a change in the object distance is manifested as a radial scaling of a PSF as also shown in Fig.~\ref{fig:intro_complete}(b).
Therefore, by designing a coded mask that has no structure in the radial direction, it is possible to construct a lensless measurement system that cutoffs depth-dependent information, resulting in extended-DOF imaging using only a single deconvolution filter.
In a previous study~\cite{nakamura_etal_radial_IAOC_2020}, this effect was verified by simulations but not by optical experiments.
As for the structure of the radial mask, only one star-chart pattern with periodicity in the rotational direction was verified; however, the mask pattern should be optimized to maximize the goodness for spatial imaging while maintaining its characteristics of depth invariance.

\subsection{Coded Mask Optimization (Related Works)}
A coded amplitude mask can be represented as a 2D tensor~$\textbf{M}\in\mathbb{R}^{N_\mathrm{y} \times N_\mathrm{x}}$ where the value of each element represents the light transmittance related to spatial coordinate~$(x, y)$ at the mask plane.
Here $N_\mathrm{y} \in \mathbb{N}$ and $N_\mathrm{x} \in \mathbb{N}$ are the vertical and horizontal pixel count of a spatially discretely represented mask. 
For the optimization of a light-transmittance pattern of a mask, i.e. the value of each element in a tensor~$\textbf{M}$, Horisaki~{\it et al.}~\cite{Horisaki_etal_DeeplyCoded_OL_2020} proposed a joint-optimization technique of the mask pattern and a reconstruction deep neural network.
This work is positioned as an application of the techniques known as end-to-end optimization~\cite{Sitzmann2018, Zhou2021} or deep optics~\cite{Hain2018,Chang2018} to the design of lensless cameras.
In the methodology, the simulation-based forward sensing model with mask variables and the reconstruction model based on a convolutional neural network~(CNN) are simultaneously optimally designed by supervised training using a large amount of prepared data pairs.

In lensless imaging, optical transmittance is preferred to be designed in binary form to facilitate the implementation of the coded mask; however, using a training algorithm based on continuous optimization, the output transmittance should be a continuous value.
Therefore, an additional technique is required to realize the binarization of mask variables.
To address this issue, Horisaki {\it et al.} simply binarized the mask variables after training~\cite{Horisaki_etal_DeeplyCoded_OL_2020}.
Instead of manual binarization, Bacca {\it et al.} also proposed the end-to-end optimization of the lensless imaging system, where they enforced mask parameters to be quantized, and these quantized values can be enforced to be $0$ and $1$~\cite{Bacca_etal_DeepCoded_IEEE_2021}.
To the best of our knowledge, the coded mask optimization processes proposed in previous studies have generated random coded masks because they do not impose any constraints on the shape in the spatial structure of the coded masks' parameters.
When considering DOF extension, it is reasonable to constrain the mask shape to be radial, based on prior knowledge of physics.

\subsection{Paper Overview}
In this work, we address the optimization of the radial mask for lensless imaging and the verification of the extended-DOF lensless imaging by optical experiments with a prototype camera. Our contributions are: (1) we propose a radial-shape-preserving optimization scheme, in order to systematically identify the best parameters considering modulation transfer function~(MTF) for a radial mask; (2) we show that our optimized radial mask achieves a more balanced frequency response considering MTF when compared to naive implementations of radial masks; (3) we quantitatively and qualitatively validated the extended DOF capabilities of the optimized radial coded mask through simulations; and (4) we built a prototype lensless camera and empirically validated the extended DOF of the radial coded mask.
\section{Radial mask optimization}
In this section, we present the radial mask optimization process. We begin in Section~\ref{subsec:RadialMaskParameterization} presenting a parameterization of the radial coded mask, in order to constrain the radial shape of the coded mask throughout the optimization process. Then section~\ref{subsec:LossFunction} presents our objective function and optimization scheme for the search for the best parameters of the radial mask. Finally, section~\ref{subsec:OptimizationResults} presents the results of the optimization process and compares the optimized mask to the original radial mask proposed in the literature. 

\begin{figure}[!t]
    \centering
    \includegraphics[width=\linewidth]{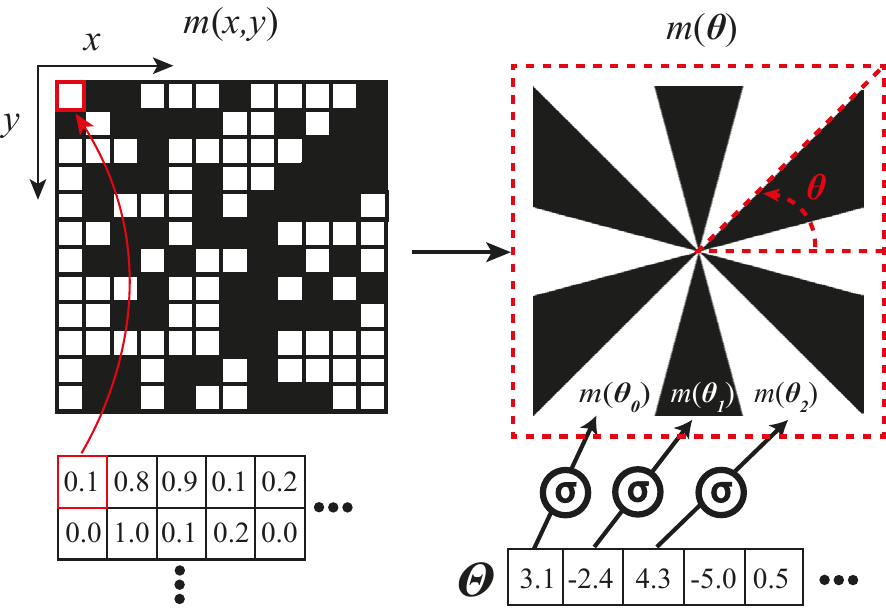}
    \caption{Overview of the radial coded mask parameterization process.
    Left: the usual modeling of a coded mask, where the light transmittance on an arbitrary discretized position in the Cartesian coordinates $(x,y)$ on the mask plane is represented as $m(x,y)$.
    Right: our proposed radial-mask parameterization, where each element of a vector $\Theta$ is related to the light transmittance~$m(\theta)$ of one of the radial sections of a coded mask. 
    The radial sections are aligned along the angular axis $\theta$.
    The logistic sigmoid function~$\sigma(\cdot)$ is multiplied on the original vector elements in order to ensure that all light transmittance values are in the range $[0,1]$.}
    \label{fig:MaskParameterization}
\end{figure}

\subsection{Radial Mask Parameterization}
\label{subsec:RadialMaskParameterization}
In past studies related to deep optics, the mask optimization process has been done independently for each element $m(x,y)$, which represents the light transmittance at the discretized Cartesian coordinates $(x,y)$ on the mask-plane.
However, this cannot be done for this work, as it is unlikely for the radial characteristic of the mask to be preserved throughout the optimization process.
In order to address this issue, we parameterize the coded mask by splitting its area into $N_\mathrm{t}$ radial sections $\boldsymbol{\Theta}=[\theta_0, \theta_1, \cdots, \theta_{N_\mathrm{t}-1}]$ along angular coordinate, where each section originates in the center of the mask and expands to its edges.
We then enforce every mask element inside the same radial section to have the same light transmittance value, as follows:
\begin{equation}
    m(x,y) = m(\theta_k)  \  \forall \ (x,y)  \ \in \ \theta_k,
\end{equation}
where $k$ is the index of the radial section, and $(x,y) \in \theta_k$ indicates a set of spatial mask coordinates that belong to the $k$-th radial section. In this scenario, the value of the light transmittance for the $k$-th radial section is:
\begin{equation}
    m(\theta_k) = \sigma( m'(\theta_k)),
\end{equation}
where $\sigma(\cdot)$ represents a logistic sigmoid function, and $m'(\theta_k)$ is a single scalar value related to the light transmittance in the $k$-th radial section. Here, the sigmoid function is employed to ensure that the values of light transmittance of the mask were in the interval $[0,1]$. For simplicity of representation, we will not explicitly write the $\sigma(\cdot)$ for the sigmoid function from this point onwards.

Here we denote a dimensionally-reduced radial-mask vector $\boldsymbol{m}_\mathrm{\theta} \in \mathbb{R}^{N_\mathrm{t} \times 1}$ that represents all the light-transmittance parameters in a radial mask along angular axis $[m(\theta_0), m(\theta_1), \cdots, m(\theta_{N_\mathrm{t}-1})]$,
and $p(\cdot)$ represents the mapping of the light transmittance elements of the radial mask in angular coordinates to a coded mask in Cartesian coordinates.
Assuming that the PSF~$\textbf{h}$ can be approximated as a mask vector itself, in which a diffraction effect is ignored, the PSF vector in Eq.~\eqref{eq:conv} can be related to the dimensionally-reduced radial-mask vector as follows:
\begin{equation}
    \textbf{h} = p(\boldsymbol{m}_{\mathrm{\theta}}).
\end{equation}


On the left side of Fig.~\ref{fig:MaskParameterization} we demonstrate a common coded mask modeling approach, where the light transmittance on a position $(x,y)$ on the plane of the mask is denoted as $m(x,y)$. On the right side of Fig.~\ref{fig:MaskParameterization}, we show our proposed radial mask parameterization process. The number of radial sections~$N_\mathrm{t}$ is a hyperparameter, and it was determined manually before the optimization experiments. We empirically selected a value of $70$ for this parameter. 
This was decided based on the resolution used for the coded mask on the optimization experiments, which was $140 \times 140$ pixels.

\subsection{Loss Function and Optimization Process}
\label{subsec:LossFunction}
For optimization, we aim at improving the frequency response, i.e. MTF, of the lensless optical system. For this paper, we desire to increase MTF values of a coded mask across all frequencies.
The MTF of a PSF~$\textbf{h}$ is defined as follows~\cite{Goodman1996}:
\begin{equation}
    \mathrm{MTF}(\textbf{h}) = \mathrm{normalize}\left(|\mathcal{F}[\textbf{h}]|\right),
\end{equation}
where $\mathrm{normalize}(\cdot)$ is the normalization operator with a value at zero frequency, $\mathcal{F[\cdot]}$ corresponds to a 2D discrete Fourier transform, and $|\cdot|$ computes the absolute values.
Our proposed loss function for mask-parameter optimization is as follows:
\begin{align}
    \mathcal{L}(\boldsymbol{m}_\mathrm{\theta}) 
    &= (-1)\times \mathrm{mean}(\mathrm{MTF}(\textbf{h}))\\
    &= (-1)\times \mathrm{mean}(\mathrm{MTF}(p(\boldsymbol{m}_\mathrm{\theta}))),
\end{align}
where $\mathrm{mean(\cdot)}$ represents the average over all elements.
Note that as we aim to increase the overall MTF of the optimized coded mask, we multiply the averaged MTF by $-1$ in order to use this loss in a minimization problem. The problem to be solved for obtaining the MTF-targeted optimized design of the radial mask can be simply written as:
\begin{equation}
\hat{\boldsymbol{m}}_\mathrm{\theta}=\argmin_{\boldsymbol{m}_\mathrm{\theta}}\mathcal{L}(\boldsymbol{m}_\mathrm{\theta}),
\label{eq:minimize}
\end{equation}
where the $\hat{\boldsymbol{m}}_\mathrm{\theta} \in \mathbb{R}^{N_\mathrm{t}\times 1}$ is the vector representing the optimized radial coded mask having dimensionally-reduced optimized parameters.
By solving Eq.~\eqref{eq:minimize} and applying the mapping function~$p$ after optimization, the radial-shape-preserved MTF-targeted optimized coded mask with spatial light-transmittance parameters can be obtained.

\begin{figure}[!t]
    \centering
    \includegraphics[width=\linewidth]{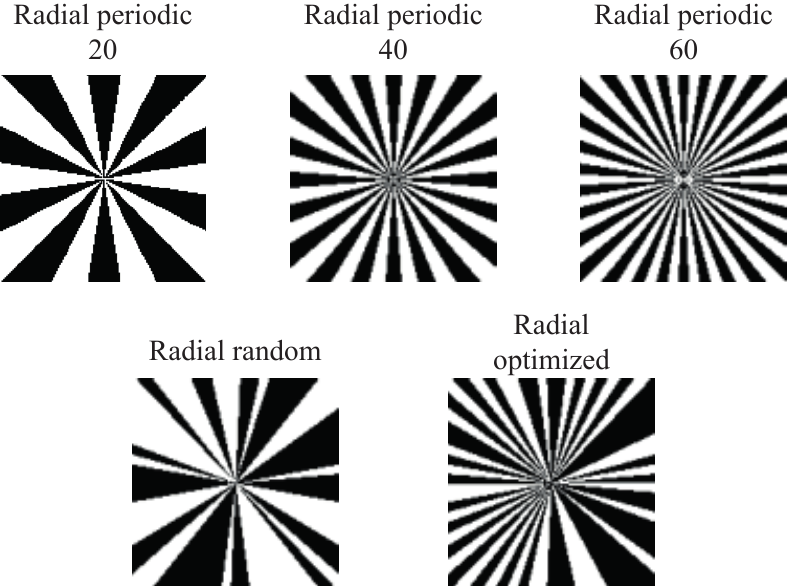}
    \caption{\red{Radial masks, including periodic masks with 20, 40 and 60 radial sections, a randomly generated radial mask, and our optimized radial mask.}}
    \label{fig:RadialMaskCompare}
\end{figure}

\begin{figure}[!t]
    \centering
    \includegraphics[width=\linewidth]{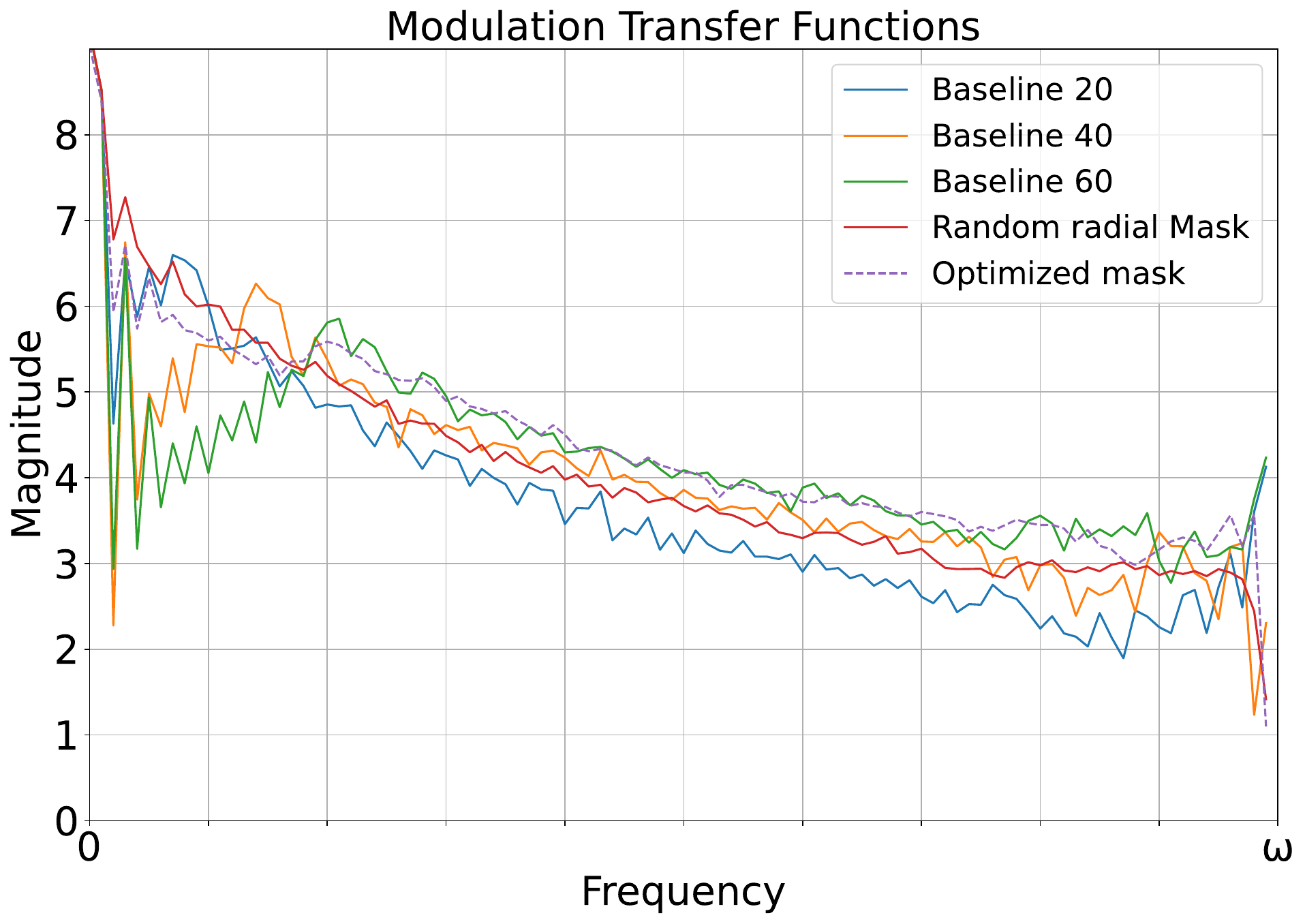}
    \caption{Comparison of MTFs of the baseline and optimized radial masks.
    $\omega$ indicates the Nyquist frequency of a mask.}
    \label{fig:MtfComparison}
\end{figure}

\subsection{Optimization Experiment}
\label{subsec:OptimizationResults}
For the optimization, we used the Adam optimizer~\cite{Kingma_Ba_Adam_ICLR_2015} with a learning rate of $0.01$ and optimized for $2000$ epochs. The values of the vector that represents the radial mask were initialized randomly with a uniform distribution inside the interval~$[-0.5, 0.5]$. As baselines for this experiment, we used the widely-used star-chart-like radial mask as Ref.~\cite{nakamura_etal_radial_IAOC_2020}.
Such a radial mask has two important properties: (1)~they are cyclical, meaning that all radial features are the same angle apart from their immediate neighbors, and (2)~they are binary in terms of light transmittance. \red{We also included a randomly generated radial mask, where we generated a radial mask with 70 radial sections, similarly to the setup of our optimized mask, but randomly assigned a binary light transmittance value to each radial section.} For the experiments, we compared our optimized mask against three baseline periodic radial masks, with 20, 40, and 60 radial sections. \red{We also included the randomly generated radial mask as a baseline.}

Figure~\ref{fig:RadialMaskCompare} presents the baseline radial masks, as well as our optimized radial mask. Similarly to the baseline masks, our optimized mask also retained a binary pattern. The average light transmittance of the optimized mask is approximately 43~\%, which is not distant to the 50~\% of the baseline hand-crafted periodic radial masks.
Interestingly, neither the binarization nor the average light transmittance were enforced explicitly throughout training and were achieved solely by optimization through our proposed MTF-targeted loss. The main difference between the optimized mask and the periodic baseline radial masks is that ours has an acyclic pattern for its radial sections.

Figure~\ref{fig:MtfComparison} presents a comparison of the MTFs of the radial masks.
For the periodic baseline masks, we observed that the cyclic characteristic of the mask defines a trade-off between low-frequency and high-frequency response. 
That is because an increase in radial sections increases the high-frequency response of the mask, but incurs a decrease of sparsity around the area of the mask that reduces the low-frequency response. \red{The randomly generated mask achieves an MTF similar to the periodic mask with 20 radial sections, where the low-frequency response is high, but at higher-frequencies the MTF is significantly lower than the periodic mask with 60 sections.} Our optimized mask, on the other hand, leveraged acyclicity to maintain areas with more and less sparsity, which improved the overall MTF values up to its Nyquist limit.
\section{Simulations}
\label{sec:simulation}
The primary goal of the radial coded mask is to extend the DOF of a lensless imaging system. So far, we have only searched for the best parameters for such a coded mask, without investigating its extended DOF properties. In this section, we determine the extended DOF of a radial mask through simulations. 

\begin{figure}[!t]
    \centering
    \includegraphics[width=\linewidth]{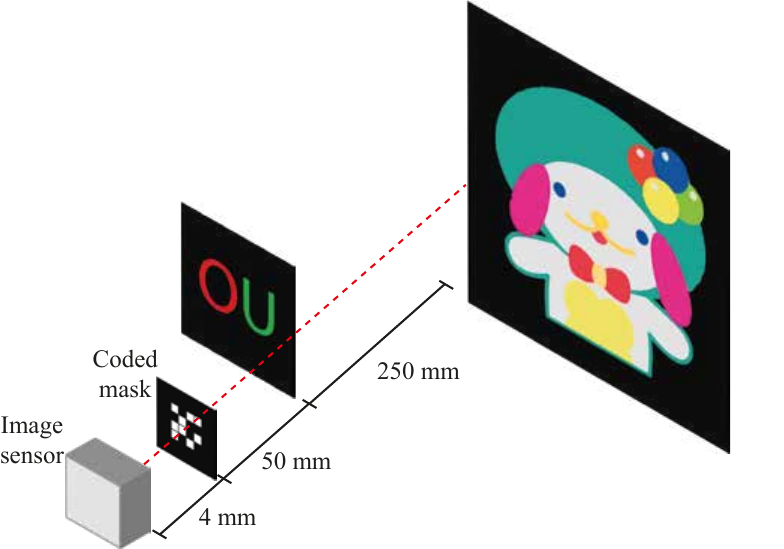}
    \caption{Geometrical setup of simulations.
    The OU pattern object was positioned at a distance of $5.0$~cm away from the coded mask, while the farther object was at $30.0$~cm away.
    The distance between the image sensor and the coded mask was set to $4.0$ mm.}
    \label{fig:SimSetup}
\end{figure}

\begin{figure*}[!t]
    \centering
    \includegraphics[width=\textwidth, trim = 0cm 0cm 0cm 0cm]{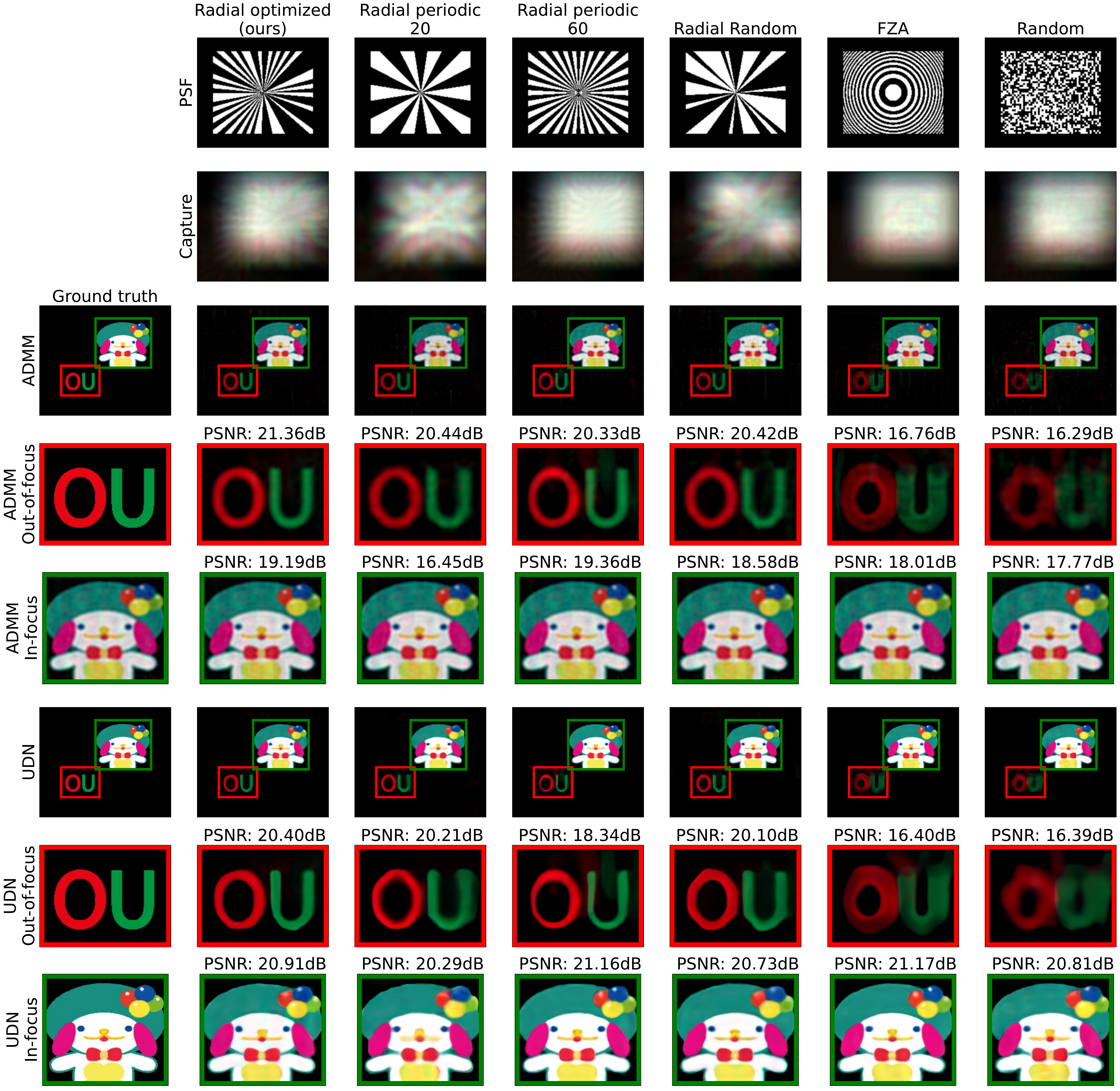}
    \caption{\red{Results of the imaging simulation.
    The top two rows present, respectively, the PSFs and sensor measurements.
    The next two rows are the reconstructed images using the ADMM algorithm and their close-ups of the OU pattern object, which cannot be correctly reconstructed by conventional methods.
    The next two rows are the reconstruction by the UDN algorithm and close-ups.
    From the left column, the results correspond to the ground truth, the radial mask, the FZA, and the random mask, respectively.}
    }
    \label{fig:SimDOF_recon}
\end{figure*}

\subsection{Conditions}
\label{subsec:DOF_sim}
The simulation was performed as a dual-depth object reconstruction, where we had two objects at two different distances from the lensless camera. We geometrically model the PSFs projected from both distances, and use them to generate the simulated sensor measurements for the full scene. The simulation of the sensing process was performed by calculating Eq.~\eqref{eq:3Dforward}. We then reconstructed an image from the full sensor measurements by using a single PSF, that could be from either of the distances. Generally, for any type of coded mask, it is expected that the object placed at the same distance at which the PSF was calibrated should be reconstructed with a higher quality than objects at different distances from the lensless camera.

We used two coded masks as non-radial baselines: a Fresnel zone aperture~(FZA)~\cite{Shimano2018, Wu2020, Nakamura2020OE} and a random mask~\cite{Nakamura2019,Zheng2021,Boominathan2020}. The random mask is a naive design for 2D lensless imaging that has good MTF up to cutoff frequency. The FZA is a coded mask with a structure only in the radial direction, opposite to the radial mask, and is suitable for digital refocusing applications. Additionally, we added two hand-crafted periodic radial masks with 20 and 60 radial sections, \red{as well as a randomly generated radial mask.}

For the lensless image reconstruction, we used two algorithms, namely the alternating direction method of multipliers (ADMM) method~\cite{boyd_etal_ADMM_NOW_2011}, and the untrained deep network (UDN) method~\cite{monakhova_etal_UDN_OptExp_2021}.
Both methods are based on the iterative error-minimization algorithm involving regularization.
For the regularization, the ADMM uses the minimization of 2D total variation~(TV)~\cite{Rubin1992} of a reconstructed image, while the UDN implements it by an untrained generative deep neural network, i.e., employment of deep image prior~\cite{DIP}.
Compared to learning-based methods~\cite{Sinha2017, Monakhova2019OE, Barba2019, Rego2022}, the results can be explainable and their precision is not restricted to a domain of learning.

Figure~\ref{fig:SimSetup} shows the simulated experimental setup.
It involves a planar plush toy and a planar OU pattern positioned $30.0~\mathrm{cm}$ and $5.0~\mathrm{cm}$ away from the coded mask, respectively.
The axial interval between the mask and an image sensor was set to $4.0~\mathrm{mm}$.
Figure~\ref{fig:SimDOF_recon} presents the mask patterns used for simulations, corresponding captured images, and reconstruction results with the ADMM and UDN algorithms using the PSF calibrated for a distance of $30.0~\mathrm{cm}$.
We set the size of the RGB captured measurements, simulated PSFs, and reconstructed images to $512 \times 612$ pixels.
In simulations, the noise was ignored to analyze the upper limit of the effect of the proposed methodology; however, a noise analysis can be drawn from the prototype camera experiments.
The coded masks used in the simulations are the same ones that were used for the prototype camera experiment.
In the reconstruction process, we used 150,000 iterations for the UDN algorithm and 100 iterations per channel for the ADMM algorithm.
The optimization code was implemented in Pytorch with a computational environment including a GPU (GeForce 3090 by NVIDIA), 32 GiB RAM, and a 10-core CPU (i9-10900K by Intel).

\subsection{Reconstruction Results}
\label{sec:Sim_results}
We limited the effective area of the mask to approximately $50~\%$ in the central region for increasing the stability of reconstruction~\cite{Antipa2018}, and the remaining perimeter of the mask was light-shielded.
From the reconstruction results of Fig.~\ref{fig:SimDOF_recon}, we observed that the plush toy was correctly reconstructed by all types of coded masks used. This was expected, as the PSF used for reconstruction was the one calibrated for the same distance as the plush-toy distance. \red{Due to higher freedom of design, we expected the random coded mask to achieve the highest PSNR for the plush toy reconstruction, which was not the case. We argue this fact is due to the presence of the OU pattern which is outside the effective DOF of the calibrated camera. In section A of the suplementary material we show that the random coded mask indeed achieves highest PSNR for the simulation when the whole scene is contained in the depth of field region for the calibrated lensless camera. More details are provided in the supplementary material. We also performed a search for the best random coded mask parameters, in order to ensure the fairness of the simulations that were performed, and the results are shown in section B of the supplementary material.}

The OU pattern, however, was placed closer to the lensless system and because of that it was better reconstructed by the radial masks. We note that the peak signal-to-noise ratio (PSNR) of the radial-mask reconstruction was significantly higher than that of the FZA and random coded masks for this object. 

\red{When comparing to the hand-crafted radial masks, we observe that the periodic masks define a tradeoff on low- and high-frequency reconstructions. More specifically, the periodic radial mask with 20 sections has a better frequency response at lower frequencies and worse response at higher frequencies, as shown in Figure~\ref{fig:MtfComparison}, and because of that it achieved a higher PSNR for the reconstruction of the sparse OU pattern object but a lower PSNR for the more detailed plush toy object. The radial periodic 60 mask achieved the opposite result, due to it having overall better MTF at higher frequencies but worse response at lower frequencies. Our optimized mask achieved a balance between these two, where the OU pattern was reconstructed with better PSNR when compared to the periodic 20 mask, while still being competitive (i.e., $\leq 0.25dB$ difference) in the plush toy reconstruction when compared to the radial periodic 60 mask. The randomly generated radial mask also achieves a tradeoff between low- and high-frequency response, however its reconstruction achieves lower PSNR for all experiments when compared to other radial masks.}

\red{In section C of the supplementary material, we show the refocusing capabilities of a lensless imaging system and also how a radial mask is capable of extending the effective depth of field of the camera independently of the distance where the PSF was calibrated from.}

\begin{figure*}[!t]
    \centering
    \includegraphics[width=\textwidth]{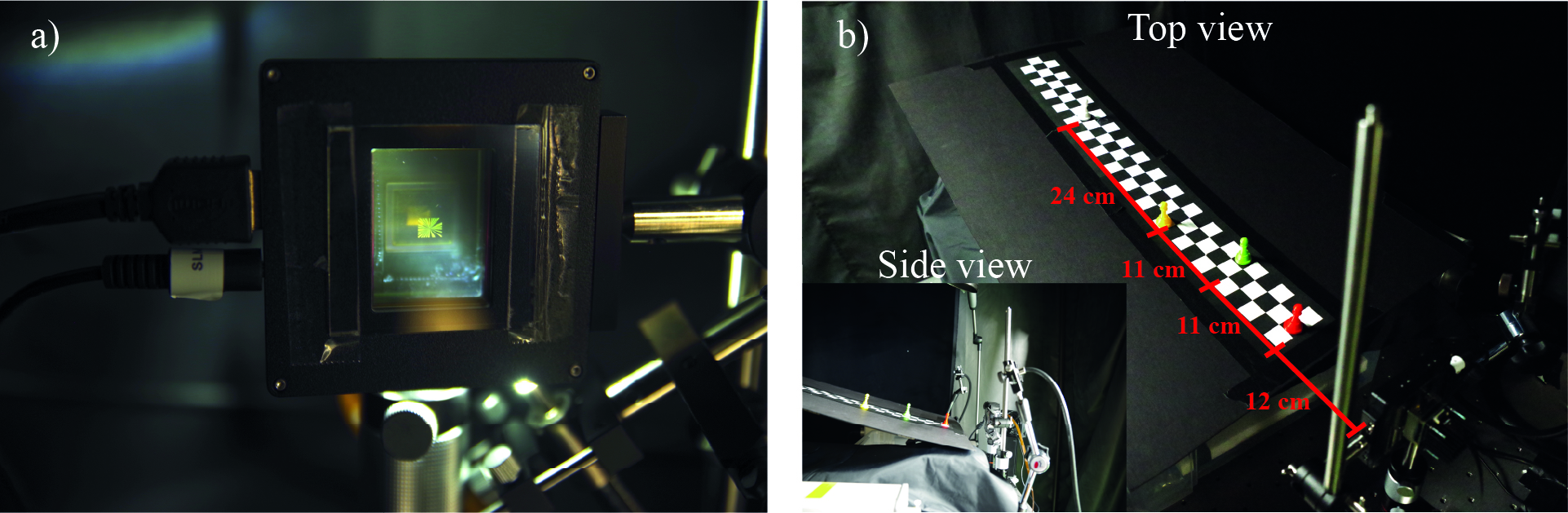}
    \caption{\red{(a)~A prototype of a lensless camera, with a spatial light modulator (SLM) creating the radial mask, and an image sensor placed behind it.
    (b)~Setup for the experiment showing the distances between coded mask and the 4 pawns. Note that the colors of the pawns in order from closer to farther from the lensless camera is red, green, yellow and white. We also present a side view of the setup for better visualization of the slanted plane containing the chessboard pattern.}}
    \label{fig:RealExpSetup}
\end{figure*}

\begin{figure*}[!t]
    \centering
    \includegraphics[width=\textwidth]{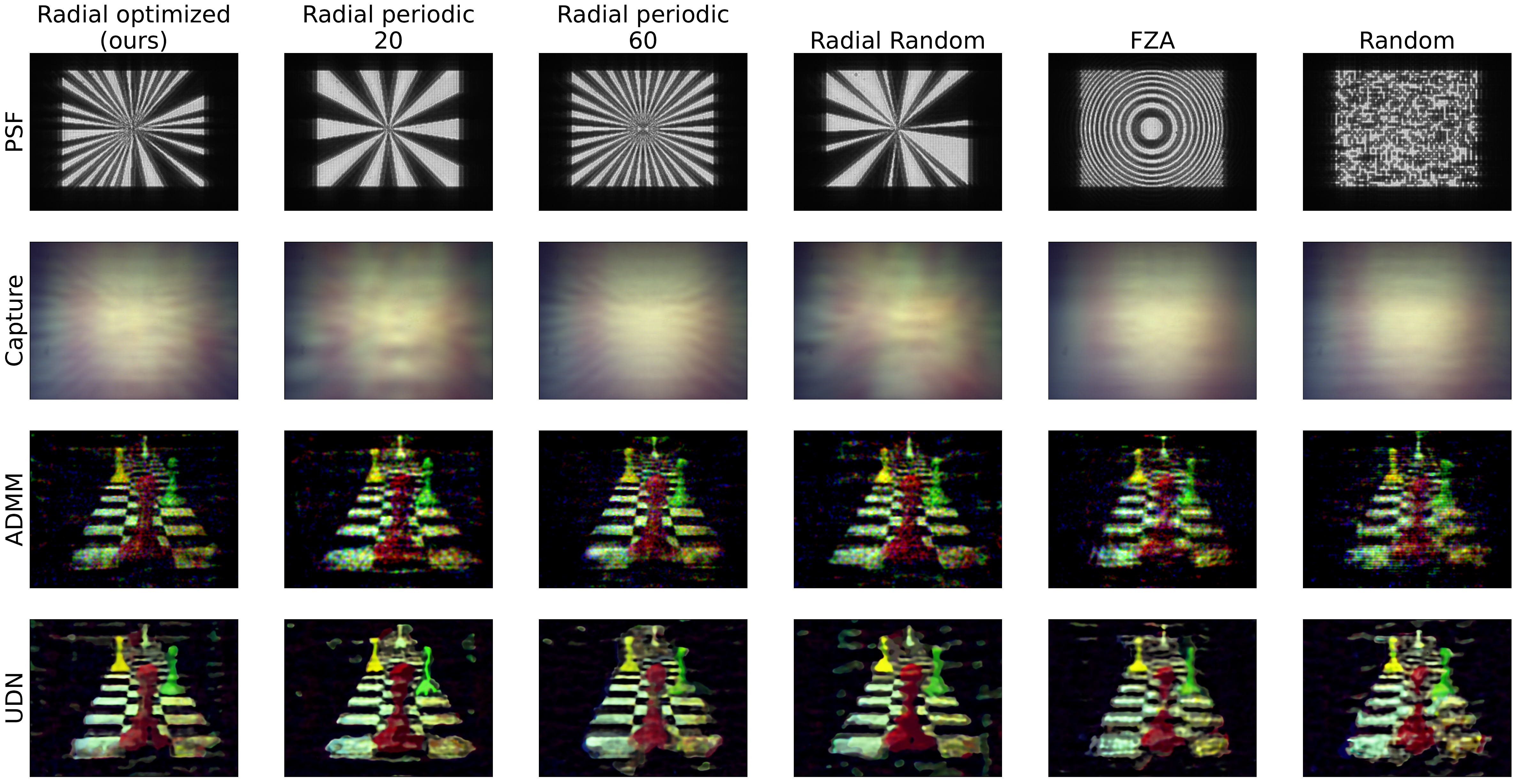}
    \caption{\red{Experimental results for a continuous depth reconstruction experiment. First row shows the calibrated PSFs. The second shows the captures from the image sensor. The third and fourth present the reconstruction results from the ADMM and UDN algorithms, respectively.}}
    \label{fig:RealExpResults}
\end{figure*}


\section{\red{Continuous Depth Optical Experiment with a Prototype Camera}}

\subsection{Setup}
\label{sec:ContinuousExpSetup}
Finally, we create a prototype lensless camera, to validate the extended DOF of the radial mask in the real world. Figure~\ref{fig:RealExpSetup}(a) shows a frontal view of the prototype lensless camera, which consists of an axial stack of a coded mask and an image sensor.
The coded mask was implemented by a transmissive liquid-crystal SLM~(LC2012 by HOLOEYE Photonics) and two polarization plates in the crossed Nicols configuration.
All coded masks to be tested are originally binary, therefore, the light transmittance of the SLM was designed as binary, and central $188 \times 228$ pixels with $36 \times 36\ \mathrm{\mu m}$ pitches were used for implementing the coded masks.
The fill factor of the SLM was $58~\%$.
Approximately $4.0~\mathrm{mm}$ behind the modulation plane of the SLM, we placed a color CMOS image sensor~(BFS-U3-28S5C-BD by Teledyne FLIR) whose pixel count was $1464 \times 1936$ with $4.5~\mathrm{\mu m} \times 4.5~\mathrm{\mu m}$ pitches.
In the experiment, $8$-bit RGB captured images were readout and they were downsampled to $732 \times 968$ pixels for reconstruction.
As well as simulations, the periphery of the mask was shielded for increasing reconstruction stability where the effective area of the mask was approximately $50~\%$.
The center of the effective area of the SLM and the image sensor were aligned by translation stages, and the planes of the two elements were adjusted to be parallel.

\red{Figure~\ref{fig:RealExpSetup}(b) shows the experimental setup including the camera and targets to be imaged.
In front of the prototype lensless camera, we placed a chessboard pattern in an inclined plane on top of which we positioned 4 pawns of different colors at different depths from the coded mask. The distances are described in Figure~\ref{fig:RealExpSetup}(b). These objects were illuminated by a white LED light installed above the lensless camera and turned towards the scene.}

\subsection{PSF Calibration}
\label{sec:PsfCalibration}
The top row in Fig.~\ref{fig:RealExpResults} shows the calibrated PSFs. The PSFs were calibrated by experimental capture of a spherical wave emitted from a light point source placed approximately $60~\mathrm{cm}$ away from the coded mask. The light-point source we used was composed of a semiconductor laser whose central wavelength was 532~nm~(Stradus 532 by Vortran Laser Technology), followed by a spatial filter~(SFB-16DMRO-OBL40-25 by SIGMA KOKI) which contained a pinhole whose diameter was $25~\mathrm{\mu m}$. The combination of the laser with the spatial filter generated a spherical wave.

\subsection{Experiments}
The second row of Fig.~\ref{fig:RealExpResults} shows the captured lensless measurements.
Although the captured image cannot be recognized by human vision, the encoded images of objects at multiple distances were multiply recorded based on the physical model in Eq.~\eqref{eq:3Dforward}. The captured image also contains color information. Note that each coded image of Fig.~\ref{fig:RealExpResults} was normalized for visualization.

\red{The PSF was calibrated by the point light source positioned $60~\mathrm{cm}$ away from the coded mask. Therefore, it is expected for the objects farther from the coded mask to be more easily reconstructed, while the closer ones should be more challenging. We observe that in both the ADMM and UDN reconstructions in Figure~\ref{fig:RealExpResults}, as the square patterns of the chessboard closer to the camera have blurry edges for the non-radial masks. Similarly, the red pawn, which is closest to the camera, is poorly reconstructed for the non-radial coded masks as well. In contrast to this, all radial masks produce an extended depth of field, being capable of reconstructing sharper square patterns. }

\red{Similarly to the simulation results presented in section~\ref{sec:Sim_results}, we can observe the tradeoff between high- and low-frequency of the periodic radial masks. The Radial periodic mask with 20 sections produces less sharp edges for the square patterns and pawns. On the other hand, the periodic mask with 60 sections produces sharper edges but noisier reconstructions. Our mask achieves a balance between these two, being capable of reconstructing sharp edges for the squares, and smooth sparse areas for the pawns.}

\section{Dual Depth Optical Experiment with a Prototype Camera}

\begin{figure*}
    \centering
    \includegraphics[width=\textwidth]{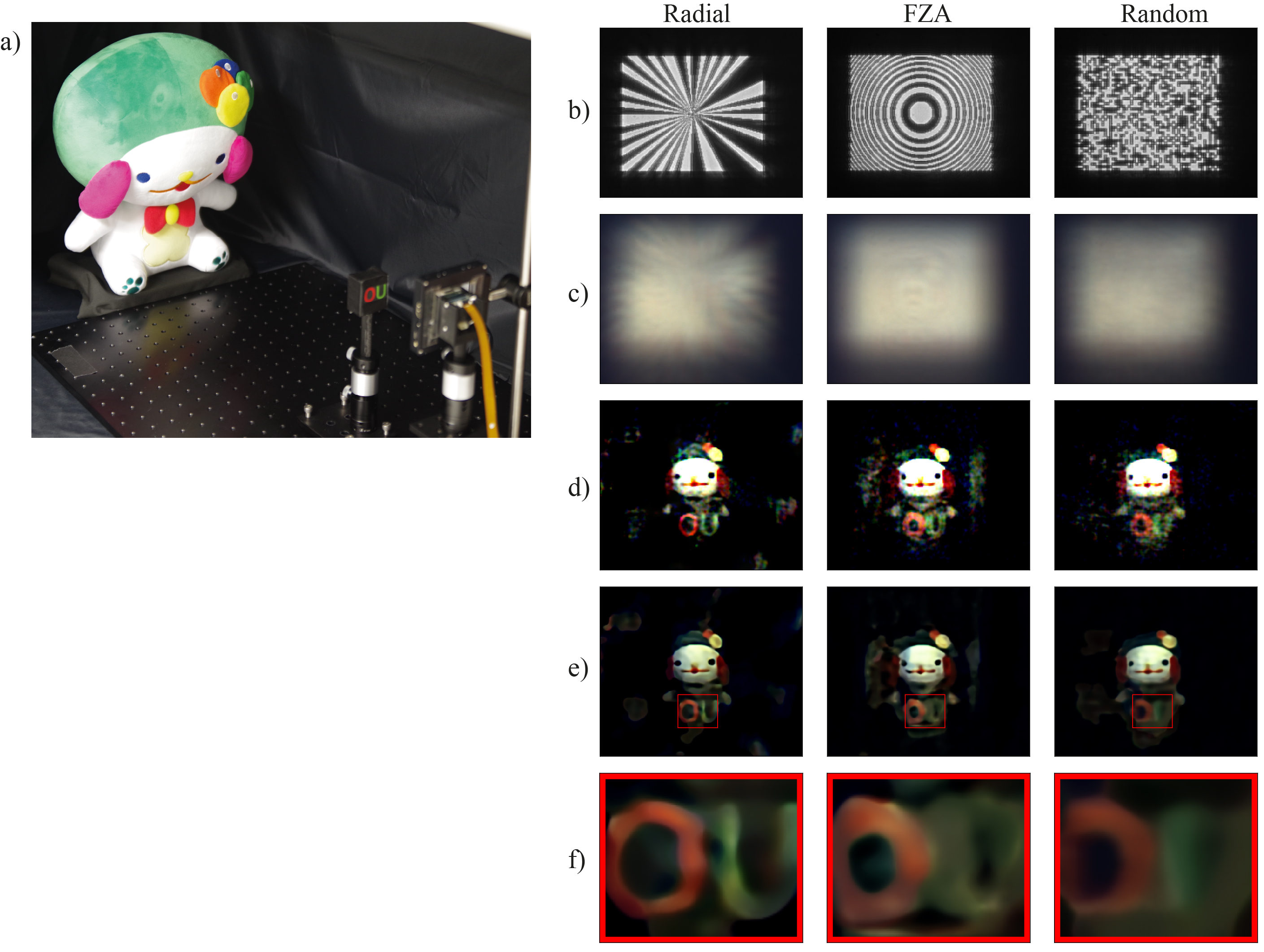}
    \caption{Dual depth object reconstruction experiment. a) Shows the experimental setup, where we placed a plush toy $30~\mathrm{cm}$ away from the coded mask, and a printed OU pattern at $9~\mathrm{cm}$ away. b) Presents the point spread functions calibrated from a distance of $30~\mathrm{cm}$ away from the camera. c) Presents the captured sensor measurements for each coded mask. d) and e) are the reconstructions by the ADMM and UDN techniques, respectively. f) Shows a zoomed in image of the printed OU pattern reconstructed by the UDN algorithm.}
    \label{fig:DualDepthExp}
\end{figure*}

We present another optical experiment, leveraging the same prototype camera presented in Section~\ref{sec:ContinuousExpSetup}, but with a simplified experimental setup to provide additional validation towards the extended depth-of-field capabilities of our proposed system.

\subsection{Setup}
The experimental setup is similar to the simulations performed in Section~\ref{sec:simulation}, in which we place two diffuse objects in front of a lensless camera. The first object is a stuffed toy known as SANKEN, which is one of the symbols of Osaka University (SANKEN plush toy). The second object is a black tape with the letters 'OU' printed on it (OU letters). The SANKEN toy and the OU letters were placed at approximately 30 and 9 centimeters (cm) away from the coded mask, respectively. These objects were illuminated by a white LED light Installed around 12 cm above the camera. 

For the calibration of the PSF to be used in the reconstruction process, we followed the setup presented in Section~\ref{sec:PsfCalibration}, but placing the light source 30 centimeters away from the coded mask.

\subsection{experiments}
Rows (b) and (c) in Fig.~\ref{fig:DualDepthExp} show the calibrated PSFs and captured measurements, respectively. Note that the captured images were normalized for visualization.

The PSF was calibrated by the point light source positioned 30 cm away from the coded mask. Therefore, it is expected that the object at a 30 cm distance to be correctly reconstructed for all three types of coded masks that were used. The OU pattern, however, was placed 9 cm away from the coded mask and was expected to be more challenging to be reconstructed properly. The coded masks and reconstruction algorithms used here were the same as those used in the simulations. The reconstructions using the ADMM and UDN algorithms are presented in rows (d) and (e) of Fig.~\ref{fig:DualDepthExp}, respectively. The bottom row (f) shows a close-up view of the OU letters reconstructed by the UDN algorithm. As expected, the plush toy was correctly reconstructed in all experiments, independently of the type of coded mask or the reconstruction algorithm employed. We note, however, that reconstructing with resolving the two lettern on the OU pattern was only successful by the radial coded mask, due to its robustness against scaling of its PSF, i.e., extended DOF characteristics. The reconstructions using the FZA and Random coded masks, on the other hand, were blurred and the two letters seemed to mix together.


\red{One alternative potential solution to the extended DOF imaging for lensless cameras is to perform monocular depth estimation\cite{Ming2021}, and subsequent use of the estimated 2 dimensional depth map to refocus the calibrated PSF so that different objects in the scene are reconstructed with PSFs calibrated for their depth. Depth estimation methods, however, are limited when facing semi-transparent and transparent planes in the ambient scene. This is caused by the fact that in such scenarios, the information from objects at multiple depths are mapped to the same pixels in the sensor and it is not obvious which depth gets estimated. We provide an additional experiment in Section D of the supplementary material to showcase the all-in-focus imaging potential of our proposed lensless camera in scenarios with semi-transparent objects where depth-based estimation techniques are expected to fail.}

\section{Conclusion}
We proposed a radial-shape-preserving optimization scheme for coded masks, which can be used to systematically create radial masks with better overall frequency response when compared to the conventional radial masks.
We showed through simulations that the optimized radial mask was capable of extending the effective DOF of a lensless camera when compared to other types of coded masks.
We also built a prototype lensless camera and empirically validated the extended DOF capabilities of the radial mask in real scenarios.

\subsection{Limitations and Future Works}
Theoretically, the PSF of a radial-shaped coded mask is depth-independent, meaning that it is not affected by radial scaling as illustrated in Figure~\ref{fig:intro_complete}(c).
That would be the case if the effective area of the coded mask were larger than that of an image sensor.
In practice, however, the coded pattern is often smaller than the effective pixel area of an image sensor, with the edges of the coded pattern being light-shielded.
This shielding is necessary when the forward model is approximately described by a convolution because it suppresses the amount of information interception by the cropping function in sensing.
In this case, even though the reconstruction processing works well and the coded pattern is depth-independent, the complete PSF formed by the coded pattern and its edge is actually depth dependent to some extent.
In future work, we will address this issue by non-convolutional, e.g. matricial, modeling of the forward problem, and/or the application of more robust compressive reconstruction algorithms such as the primal-dual splitting method, and deep-unrolling method.

In addition, this work only addressed the amplitude-modulation-type coded mask.
When compared to phase masks, one limitation of amplitude masks is the lower light-use efficiency and optical cut-off frequency.
In theory, however, the essence of this work is that the PSF is radially-shaped, and the mask implementation method and its design should be free.
Therefore, it is necessary for future works to develop masks for extended-DOF lensless imaging using phase masks with high light-utilization efficiency.

Finally, the acyclic nature of our optimized radial mask, while being leveraged for an improved frequency response, can also be a limiting factor for reconstruction of objects near the edges of the camera's field of view. That is because the low- and high-frequency components are unevenly distributed around the mask's area, and a translation of the PSF to areas near the edges of the image sensor may remove specific frequency-rich areas. Which is not the case for periodic or symmetric radial masks. 

\red{For future research, an interesting direction to be considered is to combine radial and non-radial features on a single optimized mask. In Section A of the supplementary materials, we show that a non-optimized random coded mask achieves higher in-focus PSNR for its reconstruction when compared to our optimized radial mask. An area that may be especially appealing towards combining radial and non-radial features may be polar coordinate masks\cite{Chen2017, Don2017}, that are a more generic representation of our parameterized radial mask. A promising direction would be to use a parameterized polar coordinate mask and optimize it using a loss that somehow defines a tradeoff between extended DOF and in-focus reconstruction quality.}


\vspace{11pt}
 \begin{IEEEbiography}[{\includegraphics[width=1in,height=1.25in,clip,keepaspectratio]{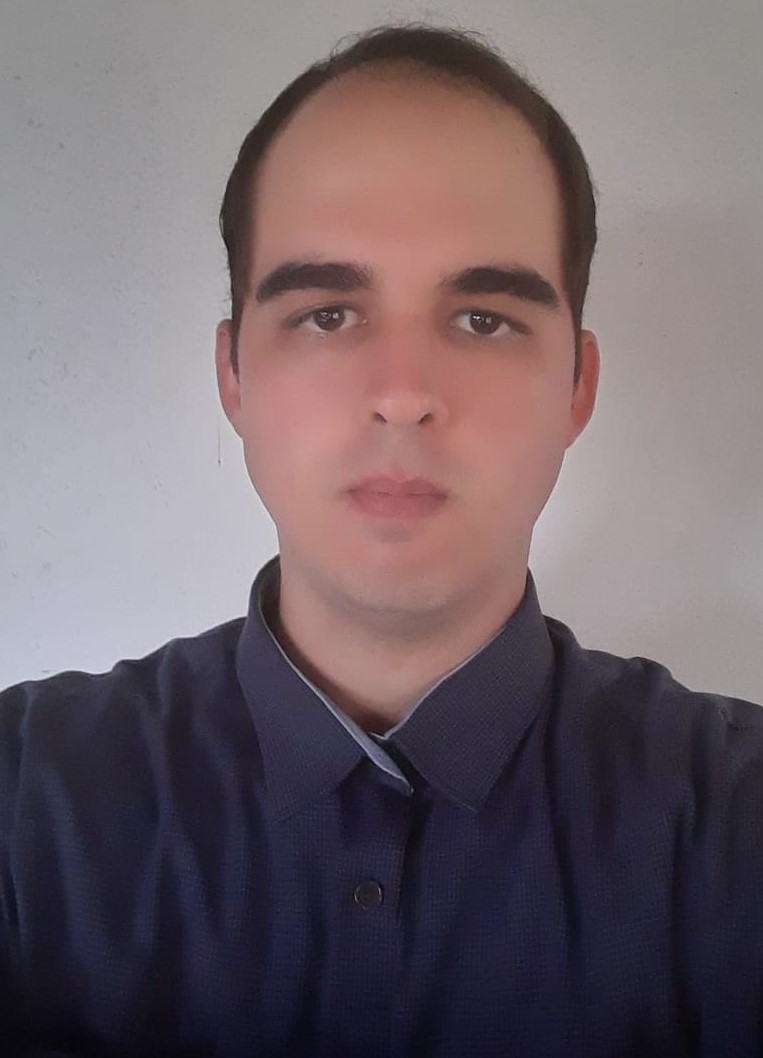}}]{Jos\'{e} Reinaldo Cunha Santos A. V. Silva Neto}
 received the B.S. in engineering and M.S. in computer science from the University of Brasilia, in 2019 and 2021 respectively. Currently, he is a PhD candidate on the computer science department of Osaka University. His research interests include computational photography, with a preference for lensless imaging, and deep learning techniques applied to computer vision.
 \end{IEEEbiography}
 \begin{IEEEbiography}[{\includegraphics[width=1in,height=1.25in,clip,keepaspectratio]{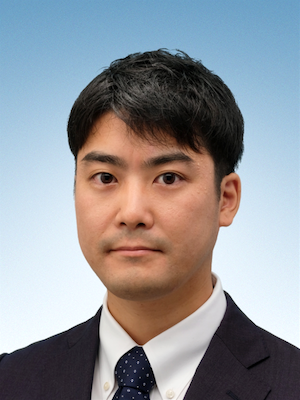}}]{Tomoya Nakamura}
received the Ph.D. degree from Osaka University, in 2015. He is currently an Associate Professor with SANKEN, Osaka University. His research interests include computational imaging, computational photography, and applied optics. He is a member of the Optica and the IPSJ. He received several honors and awards, including the International Display Workshops (IDW 2017), the Best Paper Award, and the 4th International Workshop on Image Sensor and Systems (IWISS 2018), the Open Poster Session Award 1st Place.
 \end{IEEEbiography}
  \begin{IEEEbiography}[{\includegraphics[width=1in,height=1.25in,clip,keepaspectratio]{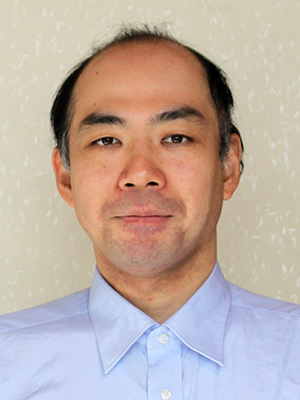}}]{Yasushi Makihara}
received the B.S., M.S., and Ph.D. degrees in engineering from Osaka University, in 2001, 2002, and 2005, respectively. He is currently a Professor with SANKEN (The Institute of Scientific and Industrial Research), Osaka University. His research interests include computer vision, pattern recognition, and image processing, including gait recognition, pedestrian detection, morphing, and temporal superresolution. He is a member of the IPSJ, the IEICE, the RSJ, and the JSME. He received several honors and awards, including the 2nd International Workshop on Biometrics and Forensics (IWBF 2014), the IAPR Best Paper Award, the 9th IAPR International Conference on Biometrics (ICB 2016), and the Honorable Mention Paper Award. He was the Program Co-Chair of the 4th Asian Conference on Pattern Recognition (ACPR 2017).
 \end{IEEEbiography}
 \begin{IEEEbiography}[{\includegraphics[width=1in,height=1.25in,clip,keepaspectratio]{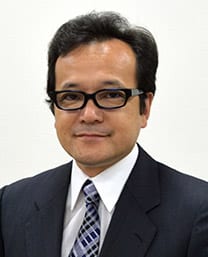}}]{Yasushi Yagi}
(Senior Member, IEEE) received the Ph.D. degree from Osaka University, in 1991. In 1985, he joined the Product Development Laboratory, Mitsubishi Electric Corporation, where he was involved in robotics and inspections. He became a Research Associate, in 1990, a Lecturer, in 1993, an Associate Professor, in 1996, and a Professor, in 2003, with Osaka University, where he was the Director of SANKEN (The Institute of Scientific and Industrial Research), from 2012 to 2015. He was the Executive Vice President of Osaka University, from 2015 to 2019. His research interests include computer vision, pattern recognition, biometrics, human sensing, medical engineering, and robotics. He is a fellow of the IPSJ and a member of the IEICE and the RSJ. He is a member of the Editorial Board of the {\it International Journal of Computer Vision}. He is the Vice President of the Asian Federation of Computer Vision Societies. He was awarded the ACM VRST2003 Honorable Mention Award, the IEEE ROBIO2006 Finalist of the T. J. Tan Best Paper in Robotics, the IEEE ICRA2008 Finalist for the Best Vision Paper, the PSIVT2010 Best Paper Award, the MIRU2008 Nagao Award, the IEEE ICCP2013 Honorable Mention Award, the MVA2013 Best Poster Award, the IWBF2014 IAPR Best Paper Award, and the {\it IPSJ Transactions on Computer Vision and Applications} Outstanding Paper Award (2011 and 2013). International conferences for which he has served as the Chair include ROBIO2006 (PC), ACCV (2007PC and 2009GC), PSVIT2009 (FC), and ACPR (2011PC, 2013GC, 2021GC, and 2023GC). He has al
so served as an Editor for the IEEE ICRA Conference Editorial Board (2008 and 2011). He was the Editor-in-Chief of the {\it IPSJ Transactions on Computer Vision and Applications}.
 \end{IEEEbiography}
\vfill

\end{document}